# *Sotto Voce:* Exploring the Interplay of Conversation and Mobile Audio Spaces

Paul M. Aoki, Rebecca E. Grinter, Amy Hurst,
Margaret H. Szymanski, James D. Thornton, and Allison Woodruff

Xerox Palo Alto Research Center
3333 Coyote Hill Road
Palo Alto, CA 94304-1314  USA

**ABSTRACT**

In addition to providing information to individual visitors, electronic guidebooks have the potential to facilitate social interaction between visitors and their companions. However, many systems impede visitor interaction.  By contrast, our electronic guidebook, *Sotto Voce*, has social interaction as a primary design goal.  The system enables visitors to share audio information – specifically, they can hear each other's guidebook activity using a technologically mediated audio eavesdropping mechanism. We conducted a study of visitors using *Sotto Voce* while touring a historic house.  The results indicate that visitors are able to use the system effectively, both as a conversational resource and as an information appliance. More surprisingly, our results suggest that the technologically mediated audio often cohered the visitors' conversation and activity to a far greater degree than audio delivered through the open air.

**Keywords**

Electronic guidebooks, shared audio, interaction analysis.

**INTRODUCTION**

A visit to a cultural heritage institution, such as a museum, is typically a social opportunity as well as an educational activity.  In fact, a shared, interactive experience with companions is often a higher priority than learning, particularly for infrequent visitors [9].   Unfortunately, many common educational tools employed by such institutions tend to reduce interaction between visitors.  For example, docent-led tours and lectures can turn visitors into a passive audience, and audio tours often isolate visitors into experiential "bubbles" [11].

Our goal is to inform the design of handheld electronic guidebooks that facilitate, rather than hinder, social interaction.  We believe that visitor engagement with co-present companions can be enhanced by providing awareness of, and context from, companions' activity. Specifically, we recommend providing direct access to the companions' guidebook audio.  This increases the resources available for engaging in conversation with companions, as well as making conversation more meaningful when it occurs.

We have reported on a previous study in which visitors could hear each other's guidebook audio through *open air*, using speakers built into the guidebook [21,22].  A key finding was that visitors oriented to the guidebooks as "participants" in a shared conversation, creating places for their turns and assigning them a role as conversational story-tellers.  However, the open air approach that enabled this shared audio experience is problematic when many visitors are present in the same location.

This paper reports on our experiences with an electronic guidebook that supports technologically mediated sharing of informational audio content.  Our design, which avoids the problem described above, is based on three key factors: headsets that do not fully occlude the ears, a careful audio design with properties that differ from those of open air, and an abstraction for audio sharing (which we call *eavesdropping*) that minimizes the interactional work needed to share.  The intimate, often directed, nature of the resulting shared audio context has led us to call the system *Sotto Voce*.

To understand the effects of our system on interaction, we conducted a study of paired visitors using the system to tour a historic house; to allow a meaningful comparison with our previous study, the study procedure remained essentially the same. We applied qualitative methods to the resulting data, including an analysis of visitor interviews and an applied conversation analytic study of recorded audiovisual observations.  The results of our study can be divided into two broad categories.  First, we found that visitors were successful in using the system.  They not only operated the device, but voiced and empirically demonstrated an understanding of the audio sharing mechanisms.  Second, a comparative analysis shows interesting changes in attentional behavior relative to [21] as well as interesting alterations in conversational behavior with respect to [22]. In half of the couples, visitors chose to use the eavesdropping feature intermittently, often in creative ways and with a social purpose. The other couples engaged in continuous *mutual eavesdropping*.  Analysis of their interactions indicated a remarkable degree of engagement and cohesion; this cohesion resulted in interactions that were "less work" and "more natural" than those found in the previous study.

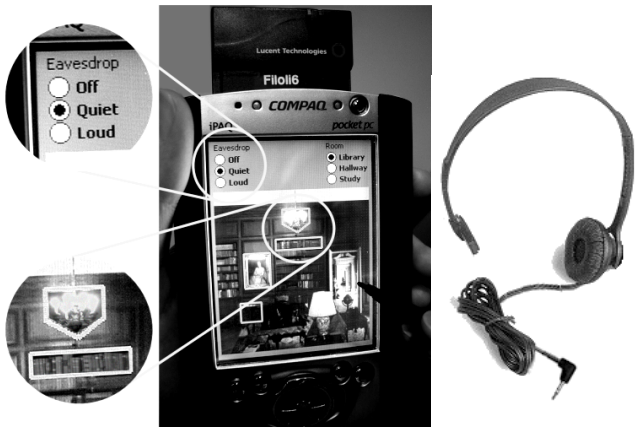

**Figure 1. Electronic guidebook and headset.**

The contributions of this paper, then, are twofold. First, we present a novel design for sharing audio information that facilitates co-present interaction. We also provide evidence of its usability. Second, we present an interaction analysis of paired visitors using the system. We anticipate that the results will be of interest to mobile audio appliance designers as well as to cultural heritage professionals.

The remainder of the paper is organized as follows. First, we discuss the design of *Sotto Voce*. Next, we describe the method employed in our user study. We then turn to findings. These are divided into general findings that apply to all visitors and more specific findings that apply to visitors who engaged in mutual eavesdropping. After discussing related work, we summarize our findings and describe future directions.

## PROTOTYPE DESIGN

Here, we concretely describe the electronic guidebook prototype used in this study. We first detail the design characteristics of the system. We then provide an expanded discussion of the rationale underlying some of the key design points. We conclude with a brief comparison of eavesdropping and sound in open air.

Our design is guided by the following principle: we want to support visitor interaction with three main entities that make demands on their attention. These entities are the information source, the visitor's companions, and the physical environment – "the guidebook, the friend, and the room" [21]. As we add capability that enhances visitor interaction with one entity, we must be careful that we do not compromise visitor interaction with other entities (e.g., we do not want to improve visitor-visitor interaction at the expense of visitor-room interaction.)

### Design characteristics of *Sotto Voce*

In this subsection, we discuss the design and implementation of the guidebook device, key aspects of its user interface, the audio sharing metaphor, and the construction of the audio content. All of these, except for the user interface, are not found in (or have features not found in) the system reported in [21,22].

*Guidebook device.* We implemented the device using the Compaq iPAQ™ 3650 handheld computer, which includes a color LCD touchscreen display. With an IEEE 802.11b wireless local-area network (WLAN) card, the device measures 163mm x 83mm x 34mm (6.4" x 3.3" x 1.3") and weighs 368g (13 oz.).

To support eavesdropping, paired devices communicate over the WLAN using Internet protocols (UDP/IP). The audio content is the same on all devices, so the devices send and receive control messages ("start playing clip 10," "stop playing clip 8") rather than waveform audio. Since our goal is to enhance co-present interaction, the device does not support remote voice communication.

*User interface.* This part of the system is very similar to that used in the previous study, and we describe its design rationale more thoroughly elsewhere [21]. Individual visitors obtain information about objects in their environment using a visual interface. This helps visitors maintain the flow of their visual task (looking at the room and its contents), which tends to reduce demands on user attention. The interface resembles a set of Web browser imagemaps; at a given time, the visitor sees a single photographic imagemap that depicts one wall of a room in the historic house (Figure 1, center). Visitors change the view perspective (i.e., display a different imagemap) by pressing a hardware button. When visitors tap on an imagemap target, the guidebook plays an audio clip that describes that object. Many, but not all, of the objects visible on the screen are targets; to help visitors identify targets, the guidebook displays *tap tips* [2] – transient target outlines that appear when the user taps and fails to "hit" a target (Figure 1, bottom left).

*Eavesdropping.* Paired visitors share audio content as follows. When visitor A selects an object on her device, she always hears her own audio clip. If A is not currently playing an audio clip, but her companion B is, then B's audio clip can be heard on A's device. In other words, audio clips are never mixed, and A's device always plays a personal clip (selected by A) in preference to an eavesdropped clip (selected by B). Audio playback on the paired devices is synchronized; if A and B are both listening to their own clips and A's clip ends first, A will then hear the remainder of B's clip as if it had "started in the middle." To control a device's eavesdropping volume (i.e., the volume at which A hears B's clips), the interface includes three option buttons: "Off," "Quiet" and "Loud" (Figure 1, top left). "Loud" is the same as the volume for personal clips.

We use commercial single-ear telephone headsets, locally modified by the removal of the boom microphone (Figure 1, right). This configuration leaves one ear available to hear sounds from the external environment.

*Audio content.* The guidebook contains descriptions of 51 objects in three rooms of the house. In most regards, the descriptions are recorded along principles described in [22]. The audio clips vary in length between 5.5 and 27 seconds, with the exception of one story that runs for 59 seconds. The clip length is much shorter than conventional audio tour clips, which often run to 180 seconds, and is intended to facilitate conversation by providing frequent opportunities for visitors to take a conversational turn.

Since we use single-ear headsets, both personal and eavesdropped audio content are necessarily presented in the same ear. We distinguish the two types of content using two mechanisms. First, we apply a small amount of reverberation to the eavesdropped audio. A single earphone cannot effectively deliver spatialized audio [4], but can support other sound effects; we chose reverberation after conducting user tests (n=6) involving scenario-based tasks using the guidebook. Second, the default eavesdropping volume ("Quiet"), which was frequently used by visitors, is softer than the personal volume.

**Audio design rationale**

Two design decisions required particular attention and experimentation. The first was the use of separate audio channels for content and conversation. The second was the abstraction, or model, we presented to the visitors for the control of audio sharing.

*Split channels.* The single-ear design described in the last subsection was not a starting assumption of our work. Our goal was to provide the following three desirable capabilities: individual control over the audio content, the ability to converse, and the ability to share content. Commercial audio tours do not address all three capabilities [11]. Playing audio content into open air supports all three capabilities, but informal experiments conducted by commercial audio guide vendors have confirmed that this approach is problematic in a public space with a large number of visitors [L. Mann, Antenna Audio, personal communication]. We therefore looked for alternatives.

We conducted user tests (n=8) of a wide variety of headsets to determine their compatibility with our design goals. We evaluated one-ear and two-ear headsets based on a variety of over-ear (ear cup, ear pad) and in-ear (ear bud, ear canal plug, ear tube) earphone designs; the in-ear designs based on ear tubes do not occlude the ear passage, enabling users to hear with both ears. Each participant was observed while performing a task involving extended attentive listening, replicated for each headset, and was then interviewed. The main parameters of inquiry were audio quality, ability to converse and comfort. Three findings determined our choice of headset. First, strong (though not always unanimous) objections on the grounds of comfort led to rejection of nearly all headsets with in-ear earphones, including the non-occluding headsets. Second, the remaining in-ear designs leaked excessive amounts of sound into the external environment. Third, the two-ear headsets had a strong isolating effect and inhibited the ability to converse. By contrast, all single-ear headsets enabled participants to converse easily.

As a result of the tests, our final design includes a headset with a single over-ear earphone (Figure 1, right). Conversation and content are therefore presented in separate channels. This design has two benefits in addition to facilitating conversation. First, *dichotic* (channel per ear) presentation is relatively effective at enabling listeners to distinguish the channels [7]. Second, ambient sound can be heard in the open ear, reducing the "bubble" effect.

*Control model.* We considered many abstractions for user control of audio sharing. We initially envisaged a simple audio space model that closely resembled "open air." However, we considered other options, such as a telephony-like connection model in which visitors would independently initiate and terminate audio sharing sessions with their companions. We also considered email-like asynchronous models in which visitors would send and receive audio clips at their convenience.

After assessing the relative demands on user attention, we returned to an audio space model. We rejected more complex abstractions that involved multiple actions (send/receive, connect/accept/reject, etc.) because we believed that the necessary interface gestures would distract visitors from their experience with the environment and their companions. In the audio space model, sharing requires no gestures of its own. To "receive," a visitor merely sets the eavesdropping volume. To "send," a visitor simply selects an object; playing a description has the side effect of sharing it, if the companion chooses to eavesdrop. The audio space model has the further advantage that it supports simultaneous listening, which enhances social interaction by creating the feeling that the content is part of a shared conversation [22].

**Comparing eavesdropping to open air**

The last two subsections have shown that eavesdropped audio, while similar to sound overheard through open air, also has important differences. The audio space model and synchronization of shared clips create effects that one would expect from open air. However, the dichotic presentation, constant amount of reverberation, and lack of sound mixing are quite unlike open air.

**METHOD**

The results presented in the remainder of the paper are based on a study performed at Filoli, a Georgian Revival historic house located in Woodside, California (http://www.filoli.org/). In this section, we describe the participants, procedure, and analysis techniques employed in our study. As previously noted, these are very similar to those employed in the previous study [21,22].

*Participants.* We recruited twelve study participants; four were PARC employees, six were loosely associated with PARC (family, friends, etc.), and two were Filoli volunteers. The participants constituted six pairs and, with the exception of the Filoli volunteers, had been asked to come with friends or relatives with whom they would normally visit a museum. Only one of the visitors was employed in a technical occupation, though seven reported prior exposure to handheld computers. The visitors ranged in age from 9 to "over 70," with five being 50 or over. Two of the couples were adult-child pairs and four were adult-adult pairs. Three of the couples were female-female pairs and three were male-female pairs. Half of the visitors described themselves as frequent museum visitors (visiting museums three or more times a year). One couple had participated in the previous study.

*Procedure.* Each pair of visitors was observed during a private tour of the house. At the beginning of the tour, each visitor was fitted with a wireless microphone. The tour consisted of three distinct phases, detailed below. The entire procedure took approximately 90 minutes; no time limits were imposed during any portion of the procedure.

In the first phase, the visitors toured eight rooms using the house's existing paper guidebook. During this phase, a member of the research team escorted the visitors to answer questions. The visitors' comments and conversation were recorded using the wireless microphones.

In the second phase, the visitors toured three rooms using the electronic guidebook. The research escort distributed guidebooks to the visitors and then gave brief instructions on the operation of the guidebook. The visitors were allowed to move through the three rooms without constraints, i.e., they were not instructed to remain together or to interact. Visitors typically spent about 20-25 minutes using the electronic guidebooks. The visitors' comments and conversation were recorded using the wireless microphones, the visitors were videotaped using a combination of manned and fixed cameras, and the visitors' use of the guidebooks was logged by the device.

The third phase consisted of a semi-structured interview conducted by two members of the research team. The interviews lasted about 30 minutes.

*Analysis.* We analyzed the data from the second and third phases using a variety of techniques. For example, we transcribed and analyzed the interview data to examine the visitors' attitudes and feelings about the technology and their experience. We also performed an interaction analysis using conversation analytic methods [15]. The interaction analysis was based on a composite video that included the audio and video recordings of the visitors, as well as the audio and screen images from each visitor's electronic guidebook (re-created from the guidebook activity logs). The interaction analysis was complemented by visualizations of the guidebook logs.

## GENERAL FINDINGS AND USABILITY RESULTS

In this section, we discuss behavior patterns that include all of the visitors. We briefly discuss the usability of the visual interface and make observations about the visitors' understanding and use of the eavesdropping feature.

All visitors were able to operate the visual interface after minimal instruction. None required coaching after the initial instruction session, though two older visitors who were unfamiliar with touchscreens continued to experiment with the interface for several minutes. This experience is similar to that reported elsewhere [2].

The use of eavesdropping was not uniform, but some patterns did emerge. One visitor turned off eavesdropping for the entire period, but all other visitors overheard at least one description from their companion's device, and all but one of these visitors used the default volume setting ("Quiet"). Three couples chose to eavesdrop on each other for essentially the entire period. In each of the other three couples, at least one member of the couple experimented with eavesdropping at some point during the audio tour.

We found that all visitors who used eavesdropping demonstrated an understanding of the audio space model in the observational data, the interview data, or both. In the observations, most used the shared audio as a conversational resource (i.e., made reference to, or conversationally reacted to, audio content). In the interviews, all but one described specific usages of the eavesdropping feature. Visitors who did *not* eavesdrop mutually were creative at adapting the eavesdropping mechanism for their own purposes. For example, the two parents each found ways to monitor their children's activity. As another example, two other visitors turned on eavesdropping to "free ride" on their companion's activity during, e.g., periods of fatigue or inactivity.

We anticipated somewhat more difficulty than we actually observed. The risk was that visitors would fail to have natural face-to-face interaction because the mediated audio content would be distracting, or otherwise "feel wrong," as a result of the differences between eavesdropped audio and sound heard in open air. This turned out not to be the case.

## MUTUAL EAVESDROPPING

We now focus on the behavior of couples that chose to use mutual eavesdropping. Compared to the previous study, in which visitors shared audio through open air [22], the behavior in the current study can be loosely summarized as more cohesive and aligned. We first characterize the behavior of the visitors. We then describe the major factors that resulted in behaviorial changes. Finally, we examine the guidebook's high-level impact on visitor experience.

Before we begin, we briefly review some of the assumptions and terminology that underlie the analysis, which is primarily informed by conversation analytic methods [15]. The fundamental assumption underlying conversation analytic research is that social interaction is organized into sequences of action, and the goal of the research is to describe this organization in its turn-by-turn, moment-by-moment unfolding. Two concepts from this discipline will be critical to our characterization: re-engagement and dis-engagement of talk, and the interactional organization known as story-telling.

When people are gathered together and involved in an activity, conversational interaction may occur, then lapse, then occur again. After a lapse, people *re-engage* turn-by-turn talk; alternatively, when people suspend turn-taking and *dis-engage* turn-by-turn talk, a lapse occurs. To accomplish states of re-engagement and dis-engagement, people draw upon a wide range of verbal and non-verbal communicative resources as well as features of the activity in which they are involved [19].

In the conversation analytic literature, *story-telling* denotes a specific, sequential organization [14]. Story-telling has a three-phase organization, each phase consisting of one or more turns. First, in the *preface*, the teller sets up the story by negotiating for an extended turn. Second, the storyteller takes the extended turn during the *telling*. In this phase, story recipients often make utterances, sometimes called *backchannel*, that encourage the teller to continue. If multiple recipients are present, *byplay* [8] between recipients can occur; byplay differs from backchannel in that it communicates content to someone other than the teller, as opposed to directing encouragement towards the teller. Third, the participants share a *response* to the story. A response may be a *receipt token* ("wow", "cool") or it may extend across multiple turns.

**Characterization of mutual eavesdropping behavior**

We found that couples engaged in story-telling behavior that was centered on the electronic guidebook descriptions. This parallels the results of the previous study of audio sharing in open air [22]. Each story-telling phase can be mapped to the visitors' actions as follows: the conversation before the guidebook description is the preface, the description and any concurrent comments from the visitors comprise the telling, and conversation after the description is the response. In this subsection, the analyses show how visitors' behavior changed in each of the three story-telling phases relative to the previous study. In particular, we will see how mutual eavesdropping mode created an ongoing assumption that the couple would continue in shared activity rather than dis-engaging and pursuing independent activity. We then describe changes in the visitors' physical mobility during these story-telling episodes.

Our analyses are based upon a collection of transcribed excerpts, of which Excerpt I is representative. Excerpt I will serve as a running example throughout this section, and Table 1 summarizes the notation used. At the beginning of the excerpt (Figure 2, left), two female visitors, F and J, have just finished listening to the description of a painting. Following their response comments, F walks over to look at a second object while J begins to play the description for a third object (Figure 2, center). Shortly thereafter, F walks back towards J (Figure 2, right) and they share another response.

*Preface.* Preface talk was generally quite abbreviated. In the previous study, couples would perform extended preface negotiations for most stories (agreeing to listen to a story together, choosing an object of mutual interest, deciding who would play the description, etc.). In the current study, lengthy negotiations almost never occurred. For example, at the start of Excerpt I, J and F are in the middle of their visit and have shared every description up to that point. Following their response discussion (ending on line 8), J chooses the library door (line 9). In this case and others, the preface was not verbalized, being implicit in the selection of the description. J verbally signals her choice of object to F (line 11) after the audio begins playing (line 10). In other words, the verbal portion of the preface overlaps with the start of the telling phase. Such reduced coordination is suggestive of ongoing activity.

A more subtle behavior also demonstrated this supposition of ongoing, shared activity. In the previous study, some participants initiated stories with questions of the form, "How about [selecting] *X* [and listening to its description]?" Such a proposal says 'I'm asking you to do this with me' and does not presuppose engagement in shared activity. By contrast, in the current study, some couples began a story-telling by asking, "What do you want to look at?" or "Which one do you want to see?" Such questions strongly imply that 'We're doing this together' and that a choice of description is being offered, as opposed to a choice of continuing the shared activity.

*Telling.* Couples frequently engaged in byplay talk. In the previous study, visitors limited themselves to backchannel-like utterances ("wow," "huh"); there were few instances of longer utterances in the telling phase. In the current study, visitors would often communicate reflective responses while the audio description was playing. For example, when listening to a description of a portrait, one visitor exclaimed, "She's pretty!" to which her companion responded jokingly, "Yeah, it was probably the painter's job!" As another example, in Excerpt I lines 17 and 19, J interacts with F regarding the guidebook's functionality as the audio description is playing.

*Response.* All couples engaged in mutual eavesdropping had some story responses that consisted of an extended conversation. In the previous study, responses were often limited to receipt tokens; extended, multi-turn conversations rarely occurred. In the current study, all of the couples engaged in response phase conversations that were more substantive than those seen in the previous study. The audio content was often the springboard for these conversations. For example, following the audio description in Excerpt I (lines 10-22), J and F reflect on the fact that "only privileged people" entered these doors (lines 23-25); J play-acts by calling for the dummyboard, a painted flat figurine that depicts a person, which they had heard about in the previous room. The conversation then continues for additional turns. In many cases, participants responded to a description with directly related questions that were reflective ("You remember that, don't you?") or content-based ("Third quarter of what century?"). In a few cases, the talk focused on points less directly related to the audio ("That reminds me of my brother's…").

*Physical interaction.* Visitors were noticeably more mobile during periods of engagement. In the previous study, the audio shared through open air was at a low volume, so couples tended to stay close together and stationary during

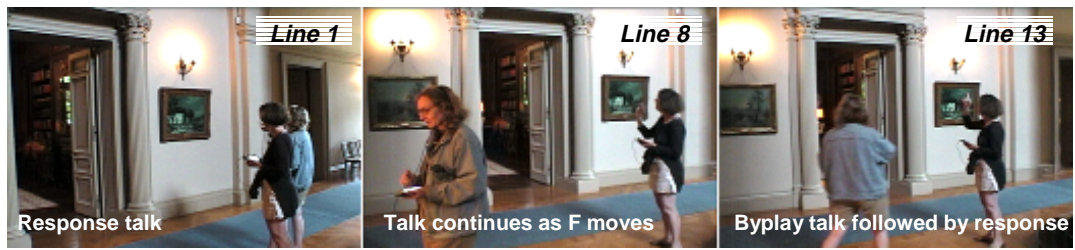

**Figure 2. Visitors interact during movement and audio descriptions (line numbers refer to Excerpt I).**

| X:        | Visitor X is speaking. **((**comment or action**))** |
| X-PDA:    | *Visitor X's guidebook is speaking.* |
| **(***n***)** | A conversational pause of *n* seconds. |
| my **[**talk<br>    **[**your talk | Alignment of overlapping speech or actions. |

**Table 1. Summary of transcription notation.**

**Excerpt I.**

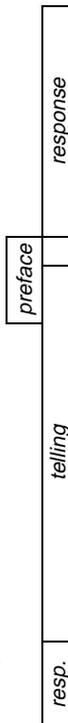

**1**  F:     I really like that one a lot.
2         (0.2) I wish they'd put the worth of some
3         of these things, ((F starts walking left of J))
4         I'd just be cur[ious,
5  J:                    [take it all to uhm
6         what's that r- antiques roadshow.
7         (0.4)
**8**  F:     oheh yeah.
9         (0.3) ((J selects Library Entrance))
10 J-PDA: *Both* [*the library and drawing room are*
11 J:           [here, ((points to doorway))
12 J-PDA: *entered through doors* [*surrounded*
**13**                                [((F walks back to J))
14 J-PDA: *by architectural* [*features*
15                           [((J points to doorway))
16 J-PDA: *like Greek columns called aedicules.*
17 J:     does yours [highlight when mine's
18 J-PDA:            [*When the house was used for*
19 J:     [((looks at F-PDA)) no.
20 J-PDA: [*entertaining, these elaborate door surrounds*
21         *communicated the fact that only privileged*
22         *people were invited to enter these rooms.*
23 J:     ((in English accent)) pull out the dummyboard,
24 F:     heh,
25 J:     ((said laughingly)) only privileged people,

shared audio experiences; physical separation implied dis-engagement. In the current study, visitors using the technologically mediated audio were less constrained. Because the audio information was continuous, visitors could separate physically from each other while remaining engaged. For example, in line 3 of Excerpt I, F begins to walk away from J even as she continues conversing. When J selects the doorway description and it has begun playing, the audio pulls F back to rejoin J (line 13).

Note that this mobility does not imply that couples separated for long periods of time. The point is that physical separation did not necessarily result in dis-engagement due to loss of the shared audio context (caused by attenuation when audio is played into open air). Also, when dis-engagement did occur, the ongoing presence of the eavesdropping channel provided resources (e.g., interest-piquing information) that prompted re-engagement. These factors reduced the "expected cost" of physical separation and often resulted in increased mobility.

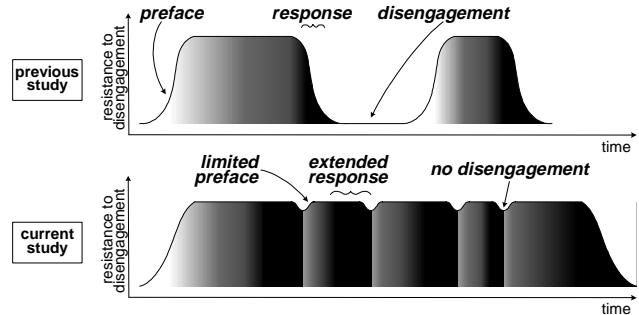

**Figure 3. Comparative patterns of engagement.**

To summarize this subsection, the couples that used mutual eavesdropping showed signs of stronger, more continuous engagement between story sequences. The typical pattern in the previous study was that the preface doubled as re-engaging talk and as the opening for a single story sequence; upon completion of response talk, the interactional state was vulnerable to dis-engagement (Figure 3, top). By contrast, in the current study, re-engagement typically occurred at the beginning of a *series* of story sequences, each separated by very limited preface talk and extended response talk. Use of mutual eavesdropping provided greater interactional cohesion and increased resistance to dis-engagement (Figure 3, bottom).

**Causes of mutual eavesdropping behavior**

The preceding characterization begs the question of causation. Fortunately, the previous and current systems are very similar, as are the study designs; as a result, there are relatively few concrete differences to which to attribute the behavioral changes. The current prototype has three main differences compared to that used in the previous study: the audio design that delivers sound through a single earphone, the "no mixing" aspect of the eavesdropping model, and the explicit availability of a sharing mechanism (as opposed to the implicit sharing in open air). For reasons of space, we limit our discussion to the key factor.

Our analysis indicates that the audio design was the most important difference. Three aspects of the audio design stand out. First, the use of earphones improves the volume and clarity of the descriptions, and the dichotal presentation allows conversation and the audio descriptions to occur "in separate ears." The audio quality is more immediate and intimate than that resulting from speakers played into open

air, as one visitor reported in the interviews:

J: [I]t definitely made it more personal between us… I could envision very easily separating, looking at different things – we did it a couple of times, I think – and still having that sense of doing something that we were doing together.

Second, recall that, unlike sound in open air, eavesdropped audio does not attenuate when visitors separate. This affects the inclination of visitors to move apart, as well as their ability to eavesdrop when they are already separated. Finally, the audio playback synchronization successfully preserved the visitor's sense of being "spoken to at the same time," which was shown in the open air study to promote co-present social interaction.

**Discussion and implications of mutual eavesdropping**

In the previous two subsections, we presented observations about behavior with technologically mediated audio relative to that exhibited in the previous study (using open air audio). In this subsection, we draw these observations together to discuss the effects of mutual eavesdropping on the visitor experience. The observations can be organized into three themes: a change in the structure of the visitors' activity (relative to the previous study with open air audio), the impact of this change on relationships between companions, and the impact of this change on relationships between visitors and their physical surroundings.

Visitor activity was structured very differently with eavesdropped audio than with open air audio. With open air audio, visitors focused on choosing individual objects and coordinating with their companions to listen to the descriptions. Repetitive setup focused more attention on the guidebook and coordination activity than seems necessary or desirable. With eavesdropped audio, the supposition of continuing shared activity meant that setup tended to be cursory, having the effect of pacing or synchronizing an ongoing activity rather than coordinating discrete acts. By reducing the effort needed to choose and listen to descriptions, mutual eavesdropping freed visitors to direct more attention to meaningful interactions with their companions and the room and its contents (i.e., away from the guidebook and routine coordination).

Couples that used mutual eavesdropping showed evidence of strengthened interactional ties. In interviews, most visitors reported a strong feeling of "connection," even while physically separated, and evidence supporting this claim recurred throughout the observational data. For example, when a visitor played a clip, the actions of both the player and the listening companion indicated that the player was accountable for subsequent actions (e.g., the listener complained when the player interrupted the clip, and the player would apologize); this kind of accountability was less in evidence in the previous study. Perhaps most convincingly, visitors participated in far more natural, rewarding forms of conversation. Visitors used casual forms of talk (e.g., byplay and extended conversations) that were not seen in the previous study, and the reduction in low-quality coordination talk meant that a higher proportion of talk tended to focus on topics of substance.

(It is worth mentioning that some couples, particularly those who did not mutually eavesdrop, talked less during their time with the electronic guidebook than with the paper guidebook, but much of the talk in the latter situation was actually coordination or conversational "filler" – visitors even self-reported this in interviews.)

Our final claim is that couples that used mutual eavesdropping exhibited an increased awareness of the room and its contents. For example, the examination of objects was more frequently occasioned by their presence in the *room* rather than their presence in the *guidebook*. In one instance, a pair of visitors looked at a set of family portraits, physically scattered around the room, as a sequence; interactions displaying this kind of orientation – i.e., at the granularity of a thematic collection rather than a single object – never occurred in the previous study. Further, visitors often examined and discussed objects that were not described in the guidebook, which was infrequent in the previous study. Such behavior strongly suggests that (some of) the attention "saved" by the altered activity structure was transferred to an increased awareness of the room and its objects.

**RELATED WORK**

Our work draws together three main areas of research. Each is quite substantial, and space limitations preclude a discussion of any but the most closely related work.

*Interaction in museum settings.* The importance of social interaction to museum visitors is well known (e.g., [9]). There are two studies of particular interest. McManus observed visitor usage of text labels; she noted that visitors were inclined to treat exhibit labels *as conversation* to which they had been party [12]. Vom Lehn *et al.* reported on an interaction analysis of museum visitors [20]. These studies focus on talk, interaction and learning in conventional environments; here, we have focused on the effects of electronic guidebooks on social interaction.

*Electronic guidebooks.* The cultural heritage community has formally studied electronic guidebooks (e.g., audio guides) for over 25 years [16]. Related work in HCI has focused on technological innovation (e.g., in location-aware computing [1,3]), and only recently have significant user studies been reported. For example, the Hyperaudio project reported the results of its user requirements studies [13], and the GUIDE project [6] conducted an evaluation that included observation, interviews and activity logging. These studies focus on system design and evaluation; here, we focus on the effects of our system on visitor interaction.

*Media and interaction.* There is an extremely rich literature on collaborative multimedia environments; of particular interest are media spaces [10]. Many of these systems have been evaluated, but most apply either ethnographic techniques (as in the Interval audio spaces [18]) or quantitative methods (as in Sellen's work on video-mediated conversation [17]) to studies of installed, workplace systems that provide shared access to human speech. In this study, we apply conversation analytic techniques to a study of a mobile, leisure-activity system that provides shared access to application content. A second body of work concerns the exploitation of human conversational protocols in the design of intelligent agent systems [5]. The work reported here and in [22] demonstrates that visitors will adapt properly-designed audio content into human-human interaction frameworks without any intelligent (adaptive) system behavior.

## CONCLUSIONS

In this paper, we have described a relatively simple, but carefully designed, audio sharing mechanism for electronic guidebooks. Eavesdropped content integrates into, rather than supplants, a visitor's conversational interactions. We have demonstrated that mutual use of this eavesdropping mechanism can actually result in a more cohesive social experience than that resulting from use of speakers in open air; the structure of the visitors' activity changed from one centered on coordination to one focused on substantive interaction. This, in turn, contributes to building stronger interactional ties between companions (encouraging more natural conversations) as well as increasing awareness of the room and its contents.

We continue to analyze the data from this study. We have applied the framework and methods of conversation analysis, but a full conversation analysis is far beyond the scope of this paper and is ongoing work. We are also preparing a discussion of the ways in which the visitors adapted our eavesdropping mechanism, particularly those who did not engage in mutual eavesdropping.

New work is addressing some of the open issues from this study. We are currently analyzing the data from a new study (using much larger numbers of visitors who were recruited on-site) to gain insight about inter- and intra-group interaction. We are also planning an experiment using bone conduction headsets that can provide binaural audio without occluding the ears.

## ACKNOWLEDGEMENTS
We are deeply grateful to Tom Rogers and Anne Taylor of Filoli Center for their assistance with this project. Shane Nye edited the audio and video data. Amy Hurst performed this work during an internship from the College of Computing, Georgia Institute of Technology.